\documentclass[preprint2]{aastex}

\shorttitle{DM Halos: Dynamical Basis of Empirical Models}
\shortauthors{Lapi \& Cavaliere}

\begin{document}

\title{Dark Matter Halos: \\The Dynamical Basis of Effective Empirical Models}
\author{A. Lapi$^{1,2}$ and A. Cavaliere$^1$}
\affil{$^1$Dept. of Physics, Univ. di Roma `Tor Vergata', Via
della Ricerca Scientifica 1, 00133 Rome, Italy\\
$^2$Astrophysics Sector, SISSA/ISAS, Via Bonomea 265, 34136
Trieste, Italy\\
\email{lapi@roma2.infn.it, cavaliere@roma2.infn.it}}

\begin{abstract}
We investigate the dynamical basis of the classic empirical models (specifically,
S\'{e}rsic-Einasto and generalized NFW) that are widely used to describe the
distributions of collisionless matter in galaxies. We submit that such a
basis is provided by our $\alpha$-profiles, shown to constitute solutions of
the Jeans dynamical equilibrium with physical boundary conditions. We show
how to set the parameters of the empirical in terms of the dynamical models;
we find the empirical models, and specifically S\'{e}rsic-Einasto, to
constitute a simple and close approximation to the dynamical models. Finally,
we discuss how these provide an useful baseline for assessing the impact of
the small-scale dynamics that may modulate the density slope in the central
galaxy regions.
\end{abstract}

\keywords{dark matter --- galaxies: halos --- galaxies: structure ---
methods: analytical}

\section{Introduction}

The classic S\'{e}rsic (1963) models met a wide and lasting success as
empirical representations of the projected ($2$-dimensional) light
distributions in spheroidal galaxies (for a review, see Kormendy et al.
2009). Einasto (1965) developed and used a similar shape to describe in
simple terms $3$-dimensional stellar mass profiles.

On the other hand, recent extensive $N$-body simulations (e.g., Navarro et
al. 2004; Merritt et al. 2005; Gao et al. 2008; Stadel et al. 2009; Navarro
et al. 2010) indicate that the S\'{e}rsic and Einasto functional forms also
provide good patterns to represent the spherically-averaged mass
distributions in dark matter (DM) halos ranging from galaxies to galaxy
clusters. These apply at levels comparable to, or even better than the
popular NFW formula (Navarro, Frenk \& White 1997).

Still, no agreed understanding is available to explain the value in both the
real and the virtual world of the S\'{e}rsic and Einasto representations (see
discussions by Graham et al. 2006; Kormendy et al. 2009). Can we identify the
underlying astrophysical basis?

\section{Empirical models}

Before addressing the issue, we note that these models belong to two main
families: generalized NFW (see Hernquist 1990; Zhao 1996; Widrow 2000;
hereafter gNFW) and S\'{e}rsic-Einasto (see Graham et al. 2006; Merritt et
al. 2006; Prugniel \& Simien 1997; hereafter SE).

\subsection{Density runs}

The density runs of the SE family may be represented in the form
\begin{equation}
\hat{\rho}(\hat{r})=\hat{r}^{-\tau}\,e^{-u\,
(\hat{r}^\eta-1)}~~~,~~~u={2-\tau\over \eta} ~.
\end{equation}
Here, quantities are normalized to their value at $r_{-2}$, the reference
radius where the logarithmic slope $\gamma\equiv -{\rm d}\log\rho/{\rm d}\log
r$ takes on the value $2$; typically, in nearby elliptical galaxies $r_{-2}$
corresponds to sizes of order $10$ kpc, a few times the half-light radius
$R_e$.

The parameters $\tau$ and $\eta$ describe the inner slope and the middle
curvature of the density run, respectively. The original Einasto profile
belongs to this family, and is obtained when $\tau=0$. Note, however, that by
deprojecting from the plane of the sky a S\'{e}rsic $2$-dimensional run
$e^{-s^{1/n}}$ with index $n\approx 3-4$ (suited for normal ellipticals, see
Kormendy et al. 2009) produces a cuspy inner run as in Eq.~(1) with $\tau
\simeq 1-1.19/2n+0.22/4n^2\approx 0.8$ significantly different from $0$ and
less than $1$, as shown by Prugniel \& Simien (1997).

On the DM side, recent simulations (see Gao et al. 2008; Stadel et al. 2009;
Navarro et al. 2010) only provide an upper bound $\tau<0.9$ for the inner
slope. When the original Einasto profile (with $\tau=0$) is adopted, the
best-fit to simulated DM halos obtains for $\eta\approx 0.2$; we will come
back to this value later on.

In turn, the density runs of the gNFW family may be written in the
form\footnote{In the literature these runs are sometimes referred to as
$\alpha\beta\gamma$-models, and equivalently defined via the parameters
$\gamma=\tau$, $\alpha=1/\eta$, $\beta=\tau+\eta\xi$.}:
\begin{equation}
\hat{\rho}(\hat{r})=\hat{r}^{-\tau}\,\left({1+w\over
1+w\,\hat{r}^\eta}\right)^{\xi}~~~,~~~w=-{2-\tau\over 2-\tau-\eta\xi} ~;
\end{equation}
the parameters $\tau$, $\eta$, and $\xi$ describe the central slope, the
middle curvature, and the outer decline of the density run, respectively.
Note that familiar empirical profiles are recovered for specific values of
the triple ($\tau$, $\eta$, $\xi$); e.g., Plummer's (1911) corresponds to
$(0,2,2.5)$, Jaffe's (1983) to $(2,1,2)$, Hernquist's (1990) to $(1,1,3)$,
and NFW to $(1,1,2)$.

\subsection{Toward a single family}

The main apparent difference between SE and gNFW is constituted by the
former's exponential decline vs. the latter's powerlaw falloff $\rho \propto
r^{-(\tau+\eta\xi)}$ for large $r$.

On the other hand, Eq.~(2) is to be considered for large values of $\xi$
anyway, since a steep density run in the halo outskirts is indicated by
observations of light distribution in spheroidal galaxies (other than cDs,
see Kormendy et al. 2009), and of DM distributions from weak lensing in
galaxies and galaxy clusters (e.g., Broadhurst et al. 2008; Oguri et al.
2009; Newman et al. 2009).

The circumstance is easily translated into the formal statement that the gNFW
family \emph{converges} to the SE for large $\xi$. This is seen on recasting
$\hat \rho\,\hat r^{\tau}$ from Eq.~(2) in exponential form, to read
\begin{equation}
e^{\xi\,\ln{[(1+w)/(1+w\,\hat r^\eta)}]}\simeq e^{\xi\,w\,(1-r^\eta)}\simeq
e^{-u\, (\hat{r}^\eta-1)}~;
\end{equation}
for approximating the middle and last terms we have used the circumstance
that $\xi\gg 1$ implies $w\gg 1$ and so $\xi\,w\simeq (2-\tau)/\eta\equiv u$
applies. Thus the two families in Eqs.~(1) and (2) actually become one in
this limit.

Thus in the following we focus mainly on the SE family, and proceed to
discuss its dynamical basis in terms of the Jeans equation.

\section{The Dynamical model}

The dynamical model of DM halos hinges upon the radial Jeans equation that
expresses the self-gravitating, equilibrium of collisionless matter (see
Binney \& Tremaine 2008). The Jeans equation reads
\begin{equation}
{1\over \rho} {\mathrm{d} (\rho\, \sigma_r^2)\over \mathrm{d}r}= -\,
{GM(<r)\over r^2}-{2\beta\,\sigma_r^2\over r}~,
\end{equation}
in terms of the density $\rho(r)$, the related cumulative mass $M(<r)\equiv
4\pi\int_0^r{\mathrm{d}x}~x^2\,\rho(x)$, and the radial velocity dispersion
$\sigma_r^2(r)$. The last term on the r.h.s. describes the effects of
anisotropic random velocities via the standard Binney (1978) parameter
$\beta\equiv 1-\sigma_\theta^2/\sigma_r^2$.

Note that the Jeans equation is designed to describe a (quasi-)static
equilibrium, away from extreme major merger events like is the case with the
Bullet Cluster (see Clowe et al. 2006). But even in relaxed conditions,
solving Jeans requires an `equation of state', i.e., a functional relation
expressing the DM pressure $\rho\,\sigma_r^2$ in terms of density (and
possibly radius) only.

\subsection{Equation of state}

In seeking for such a relation, one can make contact with the classic theory
of the non-linear collapse for DM halos in an expanding Universe; here
self-similar arguments play the role of a pivotal pattern (see Fillmore \&
Goldreich 1983; Bertschinger 1985; Taylor \& Navarro 2001). This still
applies to modern views of the halo development (e.g., Mo \& Mao 2004; Lu et
al. 2006; Li et al. 2007; Lapi \& Cavaliere 2009a,b; Fakhouri et al. 2010),
that comprise two stages: an early collapse builds up the halo main body via
a few major merger events and sets its phase-space structure by dynamical
relaxation of DM particle orbits; this tails off into a secular development
of the outskirts by smooth accretion and minor mergers.

The essence of the macroscopic equilibrium is conveyed by the self-similar
scaling $\sigma_r^2\propto M/r$ adding to the geometric relation $\rho\propto
M/r^3$. The macroscopic import of the halo phase-space structure is conveyed
by combining these two quantities into the `phase-space density'
$\rho/\sigma_r^3$, or equivalently into the functional $K(r)\equiv
\sigma_r^2/ \rho^{2/3}$ often referred to as DM `entropy' (see Bertschinger
1985; Taylor \& Navarro 2001). For the latter quantity, one easily derives
the scaling $K(r)\propto r\,M^{1/3}$ implying
\begin{equation}
K(r)\propto r^\alpha~;
\end{equation}
whence one expects a slope $\alpha$ slightly exceeding unity.

To focus the values of $\alpha$, in Lapi \& Cavaliere (2009a,b) we have
developed a full semianalytic treatment of the halo growth in the standard
accelerating Universe (see Komatsu et al. 2010). We found constant values of
$\alpha$, that fall within the narrow range $1.25-1.3$; on \emph{average},
such values grow weakly with the mass of the halo body, from galaxies to rich
clusters.

The halo development process has been probed, and the two-stage view
confirmed by intensive $N$-body simulations (e.g., Zhao et al. 2003; Wechsler
et al. 2006; Diemand et al. 2007; Hoffman et al. 2007; Schmidt et al. 2008;
Ascasibar \& Gottl\"{o}ber 2008; Vass et al. 2009; Genel et al. 2010; Wang et
al. 2010). These also confirm that: (i) a (quasi-)static macroscopic
equilibrium is attained at the end of the fast collapse, and is retained
during the subsequent stage of secular, smooth mass addition; (ii) a
persistent feature of such an equilibrium is constituted by powerlaw
correlations holding in the form $\sigma_r^2/\rho^{2/3}\propto r^\alpha$,
although it is still widely debated whether the radial or the total velocity
dispersion best applies (see also the discussions by Schmidt et al. 2008 and
by Navarro et al. 2010).

In building up our dynamical models we focus on the quantity $K\equiv
\sigma_r^2/\rho^{2/3}\propto r^\alpha$ that involves the radial dispersion
$\sigma_r^2$ (see also Dehnen \& McLaughlin 2005). Operationally, this
provides a \emph{direct} expression for the radial pressure term
$\rho\,\sigma_r^2=K\,\rho^{5/3}\propto r^\alpha\,\rho^{5/3}$ in the Jeans
Eq.~(4); anisotropies are accounted for by the last term on the r.h.s., as
discussed in \S~3.3 below.

\subsection{The DM $\alpha$-profiles}

In terms of $K(r)\propto r^\alpha$, the Jeans equation may be recast into the
compact form
\begin{equation}
\gamma = {3\over 5}\,\left(\alpha+{v_c^2\over \sigma_r^2}+2\,\beta\right)~,
\end{equation}
with $\gamma\equiv -{\rm d}\log\rho/{\rm d}\log r$ representing the
logarithmic density slope and $v_c^2\equiv G M(<r)/r$ the circular velocity.
Remarkably, by double differentiation this integro-differential equation for
$\rho(r)$ reduces to a handy $2^{\mathrm{nd}}$ order differential equation
for $\gamma$ (see Austin et al. 2005; Dehnen \& McLaughlin 2005).

Tackling first the isotropic case $\beta=0$, we recall that the solution
space of Eq.~(6) spans the range $\alpha\leq 35/27= 1.\overline{296}$: the
specific solution for the upper bound, and the behaviors of the others have
been analytically investigated by Taylor \& Navarro (2001), Austin et al.
(2005) and Dehnen \& McLaughlin (2005). In Lapi \& Cavaliere (2009a), we
explicitly derived the solutions in the full range $\alpha =
1.25-1.\overline{296}$, that are marked by a monotonically decreasing run and
satisfy physical boundary conditions: a \emph{finite} central pressure or
energy density (equivalent to a \emph{round} minimum of the gravitational
potential); a steep outer run implying a \emph{finite} and rapidly converging
(hence a definite) overall mass. We dubbed $\alpha$-\emph{profiles} these
physical solutions.

We shall use the following basic features of the latter. In the halo body at
the point $r_0$ the $\alpha$-profile is tangent to the pure powerlaw solution
$\rho\propto r^{-\gamma_0}$ of the Jeans equation; there
$v^2_c/\sigma_r^2\propto r^{2-\gamma_0/3-\alpha}$ applies, to imply from
Eq.~(6)
\begin{equation}
\gamma_0=6-3\,\alpha~.
\end{equation}
This is consistent for $\alpha= 1.25$ with the self-similar slope; as such,
it qualifies to provide a \emph{universal} middle-range slope. Note that the
point $r_0$ lies in the neighborhood of the radius $r_{-2}$ (see \S~2.1),
specifically $r_0\approx 1.74-1.51\, r_{-2}$ holds for $\alpha\approx
1.25-1.3$.

On the other hand, a monotonic density run implies the term
$v_c^2/\sigma_r^2\propto r^{2-\gamma/3-\alpha}$ to vanish at the center; this
results in an inner powerlaw $\rho\propto r^{-\gamma_a}$ with the slope
\begin{equation}
\gamma_a={3\over 5}\,\alpha~.
\end{equation}
This differs from zero as long as the entropy run grows from the center
following $K\propto r^\alpha$ with $\alpha>0$.

Finally, a finite mass implies $v_c^2/\sigma_r^2\propto
r^{-1+2\gamma/3-\alpha}$ to hold in the outskirts, so as to yield a typical
outer decline $\rho\propto r^{-\gamma_b}$ with slope
\begin{equation}
\gamma_b={3\over 2}\,(1+\alpha)~.
\end{equation}
This exceeds the value $3$, and so constitutes the hallmark of a rapidly
saturating mass; the circumstance makes less compelling here the role of a
virial boundary.

Thus, compared to NFW the inner slope of the dynamical model is considerably
flatter and the outer slope steeper; compared to the original Einasto
profile, the main difference occurs in the inner regions where the dynamical
model is (moderately) steeper.

\subsection{Anisotropy}

It is clear from Eq.~(6) that anisotropies will \emph{steepen} the density
run for positive $\beta$, and flatten it for negative $\beta$. The latter
condition is expected to prevail in the inner region, where tangential
components develop from the angular momentum barrier (Nusser 2001; Lu et al.
2006). Moving outwards, radial motions are expected to prevail, so raising
$\beta$ up to values around $0.5$ at $r\approx r_0$; outwards of this,
$\beta$ is expected to saturate or even decrease, as one enters a region
increasingly populated by DM particles on eccentric orbits with vanishing
radial dispersions at their apocenters (see Bertschinger 1985).

This view is supported by numerical simulations (see Austin et al. 2005;
Dehnen \& McLaughlin 2005; Hansen \& Moore 2006; Navarro et al. 2010), which
in detail suggest the average anisotropy-density slope relation
\begin{equation}
\beta(r)\approx \beta(0)+\beta'\,[\gamma(r)-\gamma_a]~,
\end{equation}
to hold with parameters $\beta(0)\approx -0.1$, $\beta'\approx 0.2$, and the
constraint $\beta(r)\la 0.5$. Note that at all radii the inequality
$\gamma(r)\geq 2\,\beta(r)$ is satisfied; this has been conjectured to
constitute a necessary condition for a self-consistent spherical model with
positive distribution function (see Ciotti \& Morganti 2010).

In Lapi \& Cavaliere (2009b) we extended the dynamical model to such
anisotropic conditions in the full range $\alpha\approx 1.25-1.3$, inspired
by the analysis of Dehnen \& McLaughlin (2005) for the upper bound of
$\alpha$. We note that the latter is now slightly modified to
$35/27-4\,\beta(0)/27\approx 1.31$; likewise, the point $r_0$ where
$\gamma=\gamma_0=6-3\,\alpha$ applies moves slightly inwards, so that
$r_0\approx 1.58-1.38\,r_{-2}$ now holds.

The main outcome, however, is that the density profile is somewhat
\emph{flattened} at the center relative to the isotropic case; the inner
slope now reads
\begin{equation}
\gamma_a={3\over 5}\,\alpha+{6\over 5}\,\beta(0)~.
\end{equation}
In particular, even a limited central anisotropy (corresponding to values
$\beta(0)\approx -0.1$) causes an appreciable \emph{flattening} down to
$\gamma_a\approx 0.63-0.66$ for $\alpha\approx 1.25-1.3$.

On the other hand, we stress that such small phenomenological anisotropies
near the center imply the radial $\sigma_r^2$ and the total dispersions
$\sigma^2= \sigma_r^2\,[1-2\,\beta/3]$ to be very close.

\begin{figure*}[t]
\epsscale{2}\plotone{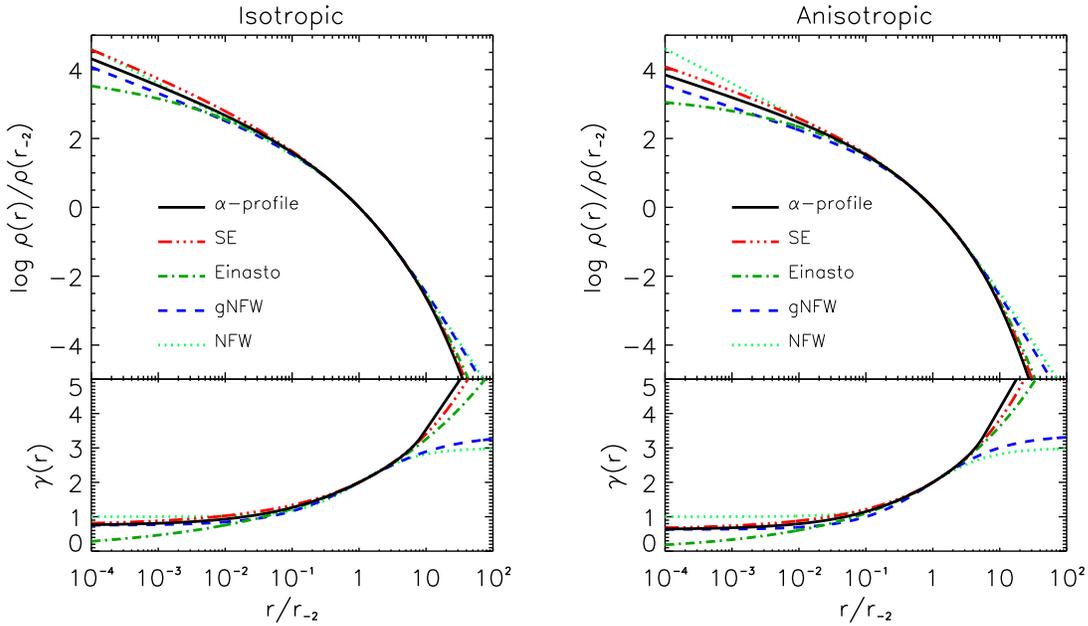}\caption{Density profiles in isotropic (left
panels) and anisotropic conditions (right panels) for the dynamical model and
its approximation in terms of empirical models with the parameters derived in
\S~4 and summarized in Table~1 and 2. The lower panels highlight the
corresponding logarithmic density slopes.}
\end{figure*}

\begin{table}[!t]
\centering\caption{Parameters of empirical models in the isotropic case}
\begin{tabular}{lccc} \hline \hline
$\alpha$ & $1.25$ & $1.27$ & $1.29$\\
\hline
\textbf{Einasto model (Eq.~1)}\\
$\eta$ & $0.211$ & $0.182$ & $0.152$\\
\\
\textbf{SE model (Eq.~1)}\\
$\tau$ & $0.750$ & $0.762$ & $0.774$\\
$\eta$ & $0.327$ & $0.287$ & $0.244$\\
\\
\textbf{gNFW model (Eq.~2)}\\
$\tau$ & $0.750$ & $0.762$ & $0.774$\\
$\eta$ & $0.687$ & $0.579$ & $0.473$\\
$\xi$  & $3.821$ & $4.564$ & $5.624$\\
\hline
\end{tabular}
\end{table}

\begin{figure*}[t]
\epsscale{2}\plotone{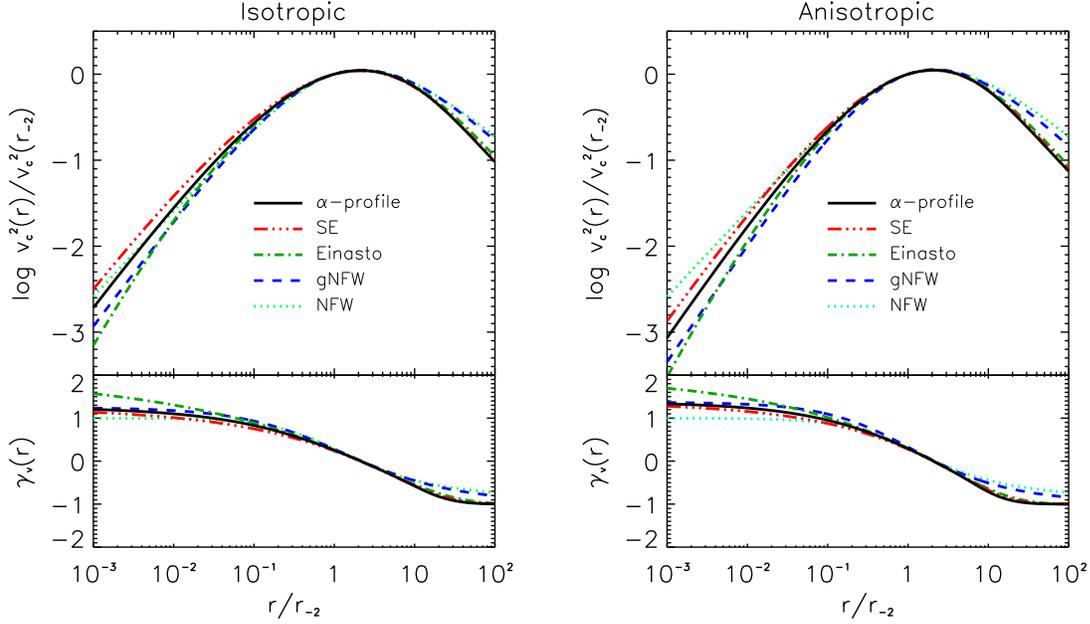}\caption{Same as Fig.~1 for the profiles of
circular velocity, and for the corresponding logarithmic slopes.}
\end{figure*}

\begin{table}[!t]
\caption{Parameters of empirical models in the anisotropic case}
\centering\begin{tabular}{lcccc} \hline \hline
$\alpha$ & $1.25$ & $1.27$ & $1.29$\\
\hline
\textbf{Einasto model (Eq.~1)}\\
$\eta$ &  $0.259$ & $0.226$ & $0.194$\\
\\
\textbf{SE model (Eq.~1)}\\
$\tau$ &  $0.630$ & $0.642$ & $0.654$\\
$\eta$ &  $0.368$ & $0.326$ & $0.285$\\
\\
\textbf{gNFW model (Eq.~2)}\\
$\tau$ &  $0.630$ & $0.642$ & $0.654$\\
$\eta$ &  $0.808$ & $0.688$ & $0.578$\\
$\xi$  &  $3.396$ & $4.018$ & $4.812$\\
\hline
\end{tabular}
\end{table}

\section{From Dynamical to Empirical Models}

Here we discuss how the parameters of the empirical profiles (see \S~2) can
be set based on our dynamical model (see \S~3); in such conditions, it will
turn out that such profiles constitute close approximations to the model over
a wide radial range.

\subsection{Parameters from dynamics}

First we consider the original Einasto profile ($\tau=0$ in Eq.~1), since
this has been widely used in the context of DM halo simulations. Here, $\tau$
is fixed to $0$, and the only free parameter is the curvature $\eta$. This we
set by requiring the logarithmic density slope
\begin{equation}
\gamma(\hat r)=2\,\hat{r}^\eta~
\end{equation}
to equal $\gamma_0$ at the point $\hat r_0$. So we find the expression
\begin{equation}
\eta  = {\log{(\gamma_0/2)}\over \log{\hat r_0}}~,
\end{equation}
that takes on values $\eta\approx 0.15-0.2$, see Table 1 and 2; remarkably,
these turn out to agree with those derived from fits of state-of-the-art
$N$-body simulations in terms of the same Einasto density run, as performed
by Navarro et al. (2010).

On the other hand, the flat central slope of the Einasto profile is at
variance with the value $\gamma_a=3\,\alpha/5$ given by our dynamical models
based on Jeans; to wit, consistency between pure Einasto and Jeans would
require at the center a flat entropy distribution and a vanishing pressure.

Actually, the simulations quoted in \S~3.1 within their finite mass
resolution provide only an upper limit $\tau<0.9$ to the central slope. This
grants scope to the full SE family of Eq.~(1).

The latter features two parameters, the inner slope $\tau$ and the middle
curvature $\eta$. These we set by requiring the logarithmic density slope
\begin{equation}
\gamma(\hat r)=\tau+(2-\tau)\,\hat{r}^\eta~
\end{equation}
to equal $\gamma_a$ for $\hat r\rightarrow 0$, and $\gamma_0$ at $\hat r_0$;
so we find
\begin{eqnarray}
\nonumber &\tau &= \gamma_a~\\
\\
\nonumber &\eta & = {\log{[(\gamma_0-\gamma_a)/(2-\gamma_a)]}\over \log{\hat
r_0}}~.
\end{eqnarray}
Thus we predict the central slope to take on values $\tau=0.6-0.8$ and the
corresponding curvature parameter to take on values $\eta=0.2-0.3$, see Table
1 and 2. It will be worth fitting the outcomes of $N$-body simulations based
on these extended SE profiles with $\tau>0$.

Finally, we report the corresponding results for the empirical gNFW family.
This features three parameters: inner slope $\tau$, middle curvature $\eta$,
and strength of the outer decline $\xi$; these we set by requiring the
logarithmic density slope
\begin{equation}
\gamma(\hat r)=\tau+\eta\xi\,{[(2-\tau)/(2-\tau-\eta\xi)]\,\hat{r}^\eta\over
[(2-\tau)/(2-\tau-\eta\xi)]\,\hat{r}^\eta-1}~
\end{equation}
to equal $\gamma_a$ for $\hat r\rightarrow 0$, $\gamma_0$ at $\hat r_0$, and
$\gamma_b$ for $r\rightarrow \infty$. So we find
\begin{eqnarray}
\nonumber &\tau &= \gamma_a~\\
\nonumber\\
\nonumber &\eta &=
{\log{[(\gamma_0-\gamma_a)\,(2-\gamma_b)/(\gamma_0-\gamma_b)\,
(2-\gamma_a)]}\over \log{\hat
r_0}}~\\
\nonumber\\
&\xi  &= {\gamma_b-\gamma_a\over \eta}~.
\end{eqnarray}

The parameters so determined are listed in Table 1 and 2 for both the
isotropic and anisotropic conditions.

\subsection{Results and comments}

With the parameters focused as discussed in the previous subsection, Fig.~1
illustrates how the empirical compare with our dynamical models. We plot the
density run of the latter (specifically for the $\alpha$-profile with
$\alpha\approx 1.25$ suitable to galactic halos), compared to those of the
Einasto, SE, and gNFW models. The left and right panels refer to isotropic
and anisotropic conditions, respectively; the popular NFW profile is also
shown for reference. To make comparisons easier, we plot in the lower panels
the corresponding logarithmic density slopes.

It turns out that the closest approximation to the dynamical model is
provided by SE, which shares with it not only the central slope by
construction, but also the body and the outer behaviors. The original Einasto
profile provides an acceptable approximation in the middle and outer ranges,
but not at the center, because of its flatness. On the other hand, the gNFW
family provides an acceptable approximation in the inner and middle ranges,
but not in the outskirts where its slope is too flat. Finally, the NFW
profile provides an acceptable approximation only in the middle range.

Similar conclusions concern the profiles of circular velocities
$v_c^2(r)\equiv G\,M(<r)/r$, that are analytically dealt with in the Appendix
and illustrated in Fig.~2.

We stress that the handy SE representation is convenient in analyzing data in
several contexts, including: the DM particle annihilation signal expected
from the Galactic Center (see Lapi et al. 2010a); rotation curves of dwarf
and normal spiral galaxies (see Salucci et al. 2007); individual and
statistical properties of elliptical and spiral galaxies (see Cook et al.
2009); strong and weak gravitational lensing (see Lapi et al. 2009b),
currently observed in clusters (e.g., Zitrin et al. 2010) and soon in massive
elliptical galaxies (see discussion by Brada\v{c} et al. 2009); X-ray
emission from the intracluster plasma (see Cavaliere et al. 2009;
Fusco-Femiano et al. 2009, Lapi et al. 2010b).

\section{Discussion}

We first stress that the dynamical model (as well as its approximations in
terms of empirical models) is in keeping with the basic features of standard
DM, i.e., its cold and collisionless nature. In fact, it implies
$\sigma_r^2(r)\rightarrow 0$ for large $r\gg r_{-2}$, a behavior expected in
the outskirts for \emph{cold} matter dominating the potential well.

At the inner end, with decreasing $r$ we expect $\sigma_r^2(r)$ to increase
toward a maximum, corresponding to effective conversion of inflow kinetic
into random energy. In fact, toward the center Jeans requires ${\rm d}\log
\sigma_r^2/{\rm d}\log r = \gamma-GM(<r)/r^2\rightarrow \gamma_a$ to hold as
the gravitational force vanishes there, to the effect that
$\sigma_r^2(r)\propto r^{\gamma_a}\rightarrow 0$.

Concerning the \emph{collisionless} nature of the DM, the boundary conditions
at the center imply a finite, non zero pressure (and energy density), while a
long collisional mean free path allows the pressure gradient ${\rm d}p/{\rm
d}r$ to diverge. Conversely, with a short mean free path $\lambda$ the
pressure gradient cannot diverge on scales $r\ga \lambda$, where a finite
$\sigma^2$ and a flatter $\gamma$ apply. In fact, weakly collisional
conditions have been proposed to explain the cored light profiles observed in
many spheroidal galaxies (see Ostriker 2000).

On approaching the center of a galactic halo, one expects the basic dynamical
model from large-scale Jeans equilibrium to be altered to an increasing
degree by \emph{small-scale} dynamics and/or energetics related to baryons.
These processes are specifically related to following issues: transfer of
energy/angular momentum from baryons to DM during galaxy formation; scouring
baryons by the energy feedback from central active galactic nuclei; any
`adiabatic' contraction of the baryons. Such issues will be briefly discussed
in turn, with a warning that they enter increasingly debated grounds.

\subsection{Energy/Angular Momentum Transfers}

Flattening of the inner density profile may be caused by transfer of energy
and/or angular momentum from the baryons to the DM during the galaxy
formation process (see El-Zant et al. 2001; Tonini et al. 2006,
Romano-D\'{i}az et al. 2008).

In detail, upon transfer of tangential random motions from the baryons to an
initially isotropic DM structure, the density in the inner region is expected
to behave as (Tonini et al. 2006)
\begin{equation}
\rho\propto
r^{-2\,[\gamma_a+2\,(2-\gamma_a)\,\beta]/[2+(2-\gamma_a)\,\beta]}~.
\end{equation}
Thus for $\beta<0$ the profile is flattened relative to the original
$\gamma_a$, down to the point of developing a core for $\beta\la
-\gamma_a/2\,(2-\gamma_a)\approx -0.3$.

However, a reliable assessment of the amount of angular momentum transferred
from the baryons to the DM is still wanting, and would require aimed
numerical simulations of better resolution than presently achieved.

\subsection{Other processes on inner scales}

Less agreed processes may affect galactic scales $r\la 10^2$ pc. For example,
at the formation of a spheroid, central starbursts and a supermassive black
hole may easily discharge enough energy ($\sim 10^{62}$ erg for a black hole
mass $M_{\bullet}\sim 10^9\, M_{\odot}$) with sufficient coupling ($\ga 1\%$)
to blow most of the gaseous baryonic mass $m\propto
r^{3-\gamma_a}/(3-\gamma_a)$ out of the gravitational potential well. This
will cause an expansion of the DM and of the stellar distributions (see Fan
et al. 2008), that flattens the central slope.

In addition, binary black hole dynamics following a substantial merger may
eject on longer timescales formed stars from radii $r\approx 10\,
(M_{\bullet}/$ $10^8\, M_{\odot})^{1/(3-\gamma_a)}$ pc containing an overall
mass of a few times the black hole's, and so may cause a light deficit in
some galaxy cores (see Merritt 2004; Lauer et al. 2007; Kormendy et al.
2009). A full discussion of the issue concerning cored vs. cusped ellipticals
is beyond the scope of the present paper.

\subsection{Adiabatic Contraction?}

On the other hand, some steepening of the inner density profile may be
induced by any `adiabatic' contraction of the diffuse star-forming baryons
into a disc-like structure, as proposed by Blumenthal et al. (1986) and Mo et
al. (1998) but currently under scrutiny, see Abadi et al. (2010).

On the basis of the standard treatments, it is easily shown that in the inner
region an initial powerlaw $\rho(r_i)\propto r_i^{-\gamma_a}$ is modified
into
\begin{equation}
\rho\propto r^{-3/(4-\gamma_a)}~;
\end{equation}
this yields typical slopes around $0.9$, steeper than the original
$\gamma_a\leq 0.78$ but still significantly flatter than $1$.

However, recent numerical simulations (see discussion by Abadi et al. 2010)
suggest that the treatment of adiabatic contraction leading to Eq.~(19) is
likely to be extreme; actually, in the inner region the contraction is
ineffective and the density slope hardly modified. Again, highly resolved
$N$-body experiments are needed to clarify the issue.

\section{Conclusions}

We have discussed the \emph{dynamical} basis of the S\'{e}rsic-Einasto
\emph{empirical} models, in terms of well-behaved solutions of the Jeans
equation with physical boundary conditions comprising: a finite central
energy density, a closely self-similar body, a finite (definite) overall
mass.

We find the SE profile to be particularly suitable to represent the general
run of the dynamical solution. Specifically, we have discussed how to tune
the parameters of SE in terms of the dynamical model; in such conditions, we
find the former to constitute a simple and \emph{close} approximation to the
latter.

The resulting SE profile shares with the dynamical model the following
features: an outer steep decline, hence a definite overall mass; a closely
self-similar body with slope $\gamma_0=6-3\alpha$; an inner slope around
$\gamma_a=3\alpha/5$, hence flatter than $-1$. The latter slope provides an
useful \emph{baseline} for discussing alterations of the inner behavior
caused by additional baryonic processes.

In conclusion, we submit that the dynamical models discussed here, namely the
$\alpha$-profiles, provide the \emph{astrophysical} basis for understanding
the empirical success of the SE profiles in fitting the real and the virtual
observables, from galaxies to galaxy clusters.

\section*{Acknowledgments}
Work supported by ASI and INAF. It is a pleasure to acknowledge inspiring
correspondence with A. Graham, stimulating discussions with L. Danese, P.
Salucci, and Y. Rephaeli, and helpful comments by our referee. A.L. thanks
SISSA and INAF-OATS for warm hospitality.

\begin{appendix}

\section*{Empirical models: circular velocities}

Here we provide analytic formulae for the circular velocity profiles related
to the empirical models presented in \S~1 of the main text. The circular
velocity $v_c^2\equiv G M(<r)/r$ constitutes a quantity helpful not only to
evaluate the gravitational potential but also in the specific context of
galactic rotation curves (e.g., Salucci et al. 2007, and references therein).

For the SE family this comes to
\begin{equation}
\hat{v}_c^2(\hat{r})=\hat{r}^{-1}\,{\Gamma\left[{(3-\tau)\over
\eta};u\,\hat{r}^\eta\right]\over\Gamma\left[{(3-\tau)\over \eta};u\right]}
\end{equation}
in terms of the (lower) incomplete gamma function
$\Gamma[a;x]\equiv\int_0^x{\rm d}t\, t^{a-1}\,e^{-t}$. On recalling that
$\Gamma[a;x]\simeq x^a/a$ holds for $x\ll 1$ and $\Gamma[a;x]\simeq
\Gamma[a]$ holds for $x\gg 1$, one finds the asymptotic behaviors $\hat
v_c^2(r)\propto \hat r^{2-\tau}$ toward the center and $\hat v_c^2(r)\propto
\hat r^{-1}$ toward the outskirts; hence a maximum occurs at a radius
$r\simeq r_{-2}$.

On the other hand, for the gNFW family the circular velocity run comes to
\begin{equation}
\hat{v}_c^2(\hat{r}) = \hat{r}^{2-\tau}\,
{_2F_1\left[{(3-\tau)\over\eta},\xi,1+{(3-\tau)\over
\eta};-w\,\hat{r}^\eta\right]\over
_2F_1\left[{(3-\tau)\over \eta},\xi,1+{(3-\tau)\over \eta};-w\right]}~
\end{equation}
in terms of hypergeometric functions $_2F_1[a,b,c;x]$. From their asymptotic
(details may be found in Abramowitz \& Stegun 1972) one again finds behaviors
similar to the SE family.

The full expressions are used to compute the profiles illustrated in Fig.~2.

\end{appendix}

\end{document}